\documentclass[twocolumn,amsmath,amssymb,nofootinbib]{revtex4}

\usepackage{graphicx}
\usepackage{dcolumn}
\usepackage{bm,psfrag}

\newcommand{\be}{\begin{equation}}
\newcommand{\ee}{\end{equation}}
\newcommand{\bea}{\begin{eqnarray}}
\newcommand{\eea}{\end{eqnarray}}
\newcommand{\ba}{\begin{eqnarray}}
\newcommand{\ea}{\end{eqnarray}}

\newcommand{\beq}{\begin{equation}}
\newcommand{\eeq}{\end{equation}}
\newcommand{\beqa}{\begin{eqnarray}}
\newcommand{\eeqa}{\end{eqnarray}}
\newcommand{\beqar}{\begin{eqnarray*}}
\newcommand{\eeqar}{\end{eqnarray*}}

\newcommand{\reef}[1]{(\ref{#1})}

\newcommand{\eg}{{\it e.g.,}\ }
\newcommand{\ie}{{\it i.e.,}\ }

\newcommand{\eqlabel}[1]{\label{#1}}  

\def\t6 {T_\mt{D6}}

\newcommand{\mt}[1]{\textrm{\tiny #1}}

\def\cala         {{\cal A}}

\def\cale         {{\cal E}}

\def\call         {{\cal L}}

\def\calo         {{\cal O}}

\def\del          {\partial}

\def\ee           {{\rm e}}

\def\sqr#1#2{{\vcenter{\vbox{\hrule height.#2pt
 \hbox{\vrule width.#2pt height#1pt \kern#1pt
 \vrule width.#2pt}\hrule height.#2pt}}}}

\def\ee{\cale}

\def\aa1{\phi}
\def\cc1{\psi}

\def\D{\Delta}

\def\lam{\lambda}

\def\Dt{\delta t}
\def\Dp{\delta p}

\usepackage[bookmarks=false]{hyperref} 
\hypersetup{pdfstartview=FitH,pdfhighlight=/O,colorlinks=false}

\begin{document}

\preprint{arXiv:1307.nnnn [hep-th]; UWO-TH-13/xx}

\title{Universality of Abrupt Holographic Quenches}

\author{Alex Buchel,$^{\! 1,2}$  Robert C. Myers$^2$
and Anton van Niekerk$^{2,3}$}
\affiliation{$^1$\,Department of Applied Mathematics, University of Western
Ontario, London, Ontario N6A 5B7 Canada\\
$^2$\,Perimeter Institute for Theoretical Physics, Waterloo, Ontario N2L 2Y5,
Canada\\
$^3$\,Department of Physics \& Astronomy, 
University of Waterloo, Waterloo, Ontario N2L 3G1, Canada}


\begin{abstract}
We make an analytic investigation of rapid quenches of relevant operators in
$d$-dimensional holographic CFT's, which admit a dual gravity description. We
uncover a universal scaling behaviour in the response of the system, which
depends only on the conformal dimension of the quenched operator in the
vicinity of the ultraviolet fixed point of the theory. Unless the amplitude of
the quench is scaled appropriately, the work done on a system during the quench
diverges in the limit of abrupt quenches for operators with dimension $\frac d2
\le\Delta < d$.
\end{abstract}

\maketitle

Quantum quenches have recently become accessible in laboratory experiments
\cite{more}, which has initiated much activity by theoretical physicists to
understand such systems. Up until now, most analytic work on the topic of
relativistic  quantum quenches have assumed that the field theory is at weak
coupling \cite{cc2}--\cite{cc4}.

The study of quantum quenches at strong coupling is accessible through the
gauge/gravity duality \cite{m1}.  Much related work studying thermalization in
the boundary theory was done by studying the gravity dual under the assumption
that the non-equilibrium evolution can be approximated by a uniformly evolving
spacetime, \eg \cite{early2}--\cite{new2}. Other approaches study the evolution
of a probe on the static spacetime \cite{das1}. The approach of numerically
evolving the dual gravity theory was initiated in \cite{cy1}. Further numerical
studies of quenches in a variety of holographic systems were presented in
\cite{adsnumerics3}--\cite{blmv}.

In \cite{blm,blmv}, holography was applied to study quenches of the coupling to
a relevant scalar operator in the boundary theory. A numerical approach was
taken to study the evolution of the dual scalar field in the bulk spacetime.
For fast quenches, evidence was found for a universal scaling of the
expectation value of the boundary operator. Similar scaling was observed for
the change in energy density, pressure and entropy density. However, no
analytic understanding of this behaviour was available.

In this Letter, we investigate these holographic quenches analytically,
focusing on the work done by the quench. Unlike \cite{blm,blmv} in which the
coupling was an analytic function of time, we abruptly (but with some degree of
smoothness) switch on this source at $t=0$. The coupling is then varied over a
finite interval $\Dt$ and is held constant afterwards. We find that for fast
quenches, the essential physics can be extracted by solving the linearized
scalar field equation in the asymptotic AdS geometry. Note that our analysis is
naturally driven to this regime by the limit $\Dt\to0$. In contrast to
\cite{blm,blmv}, we are {\it not} a priori limiting our study to a perturbative
expansion in the amplitude of the bulk scalar. Our analytic results also cover
any spacetime dimension $d$ for the boundary theory, whereas \cite{blm,blmv}
were limited to $d=4$.

\begin{figure}[h]
\begin{center}
\psfrag{ar}[c]{{$\hat{\rho}$}}
\psfrag{at}[B][tl]{{$\hat{t}$}}
 \includegraphics[width=2.8in]{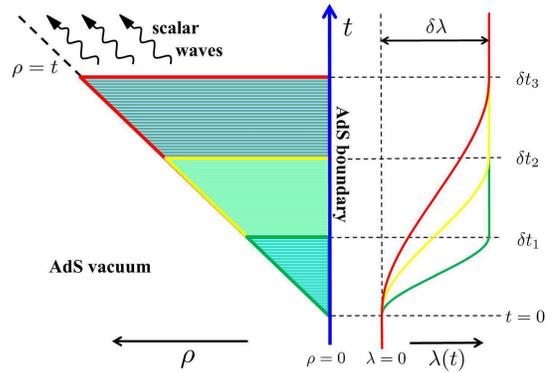}
\end{center}
  \caption{
(Colour online) The shaded triangle is the region close to the boundary of the
AdS spacetime where we must solve for the scalar field. We show several cases
with $\Dt_1<\Dt_2<\Dt_3$. The profile $\lambda(t/\Dt)$ is held fixed in each
case. In particular, the amplitude $\delta\lambda$ of the quench remains
constant as $\Dt$ becomes smaller. As the quench becomes more rapid, the bulk
region shrinks closer to the asymptotic boundary.}
 \label{triangle}
\end{figure}

Let us describe the quenches in more detail: The coupling in the boundary
theory is determined by the leading non-normalizable mode of the bulk scalar
\cite{m1}. We set this mode to zero before $t=0$, vary it in the interval
$0<t<\Dt$ and hold it fixed afterwards. Because the energy density can only
change while the coupling is changing, we are only interested in the response
of the scalar field during the timespan $0<t<\Dt$. Further, since the response
propagates in from the boundary of the spacetime, the field will only be
nonzero within the lightcone $t=\rho$. Hence to determine the work done, we
need only solve for the bulk evolution in the triangular region bounded by this
lightcone, the surface $t=\Dt$ and the AdS boundary, as shown in
fig.~\ref{triangle}. As is also illustrated, as $\Dt\to0$, this triangle
shrinks to a small region in the asymptotic spacetime. The normalizable
component of the scalar field, which determines the expectation value of the
boundary operator, can be solved analytically in this situation, and its
scaling with $\Dt$ can readily be seen from this solution.  From this, we also
obtain the scaling of the energy density in the boundary.

Consider  a generic deformation of a conformal field theory (CFT) in $d$
spacetime dimensions by the time-dependent coupling $\lam=\lam(t)$ of a
relevant operator $\calo_{\D}$ of dimension $\D$: $\call_0\ \to\
\call=\call_0+\lam \calo_{\D}$. The gravity dual describing such a deformation
is given by
\begin{align}
I_{d+1}&=\frac{1}{16\pi G_{d+1}}\int d^{d+1}x  \sqrt{-g}\nonumber\\
&\times\biggl(R+d(d-1)-\frac 12(\del\phi)^2
-\frac 12 m^2\phi^2-u(\phi)\biggr)\,,
\eqlabel{action5}
\end{align}
where we have chosen an AdS radius of $1$. The bulk scalar $\phi$ is dual to
$\mathcal{O}_{\D}$ with $m^2=\Delta (\Delta-d)$. The potential $u(\phi)$
contains terms of order $\phi^3$ or higher. To simplify our discussion, we will
consider quenches where the conformal dimension of the operator is non-integer
(for even $d$ and not half-integer for odd $d$ --- see comments below).
Further, we initially consider dimensions in the range
$\frac{d}{2}\le\Delta<d$.

Since we are interested in quenches that are homogeneous and isotropic in the
spatial boundary directions, we assume that both the background metric and the
scalar field depends only on a radial coordinate $\rho$ and a time $t$. We will
work in a spacetime  asymptotic to the AdS Poincar\'{e} patch as $\rho\to 0$.
Hence the bulk metric is
\begin{equation}
ds^{2} = -A(t,\rho)dt^{2} + \Sigma(t,\rho)^{2}d\vec{x}^2 + \rho^{-4}A(t,\rho)^{-1}d\rho^{2}.
\end{equation}
The (nonlinear) Einstein equations and the scalar field equation then take the
form:
\begin{widetext}
\begin{eqnarray}
0&=&-\frac{2(d-3)}{(d-1)A}u(\phi)+\frac{2d(d-3)}{A}+\rho^{4}\left(\phi'\right)^2-\frac{d-3}{(d-1)A}m^2\phi^2 -\left(\frac{\dot{\phi}}{A}\right)^2
+2(d-2)(d-1)\left[\left(\frac{\dot{\Sigma}}{A\Sigma}\right)^2-\left(\frac{\rho^2\Sigma'}{\Sigma}\right)^2\right]\nonumber\\
&&+\frac{2\rho^2\left(\rho^2 A'\right)'}{A}
-4\left(\frac{\dot{A}}{A^2}\right)^2+2\frac{\ddot{A}}{A^3}\,,\eqlabel{eoms1}\\
0&=&d-\frac{u(\phi)}{(d-1)}-\frac{m^2\phi^2}{2(d-1)}+\frac{\rho^4A}{2(d-1)}\left(\phi'\right)^2
+\frac{\dot{\phi}^2}{A}-\rho^4\frac{ A'\Sigma'}{\Sigma}-
(d-2)\rho^4 A\frac{(\Sigma')^2}{\Sigma^2}+
\frac{2\ddot{\Sigma}}{A\Sigma}-\frac{\dot{A}\dot{\Sigma}}{A^2\Sigma}+
(d-2)\frac{\dot{\Sigma}^2}{A\Sigma^2}\,,\eqlabel{eoms2}\\
0&=&\frac{\left(\phi'\right)^2}{2(d-1)}+\frac{1}{2(d-1)}\left(\frac{\dot{\phi}}{\rho^2 A}\right)^2
+\frac{\Sigma''}{\Sigma}+
\frac{2\Sigma'}{\rho\Sigma}+
\frac{\ddot{\Sigma}}{\rho^4 A^2\Sigma}\,,\eqlabel{eoms3}\\
0&=&\frac{\phi'\dot{\phi}}{d-1}+\frac{\dot{A}\Sigma'}{A\Sigma}-
\frac{A'\dot{\Sigma}}{A\Sigma}+2\frac{\dot{\Sigma}'}{\Sigma}\,,\eqlabel{eoms4}\\
0&=&-\frac{\delta u(\phi)}{\delta\,\phi}-m^{2} \phi+ \rho ^4 A\phi''+2\rho ^3 A \phi' +
\rho ^4 A'\phi'+\frac{(d-1) \rho ^4 A \Sigma'\phi' }{\Sigma }
+ \frac{\dot{A} \dot{\phi}}{A^2}-\frac{(d-1) \dot{\Sigma}  \dot{\phi}}{A \Sigma }
-\frac{\ddot{\phi}}{A}\,. \eqlabel{kg}
\end{eqnarray}
\end{widetext}
where dots and primes denote derivatives with respect to $t$ and $\rho$,
respectively. 
The scalar field will have an asymptotic expansion of the form
\begin{equation} \eqlabel{bc}
\phi(t,\rho)\sim\rho^{d-\Delta}\left(p_{0}(t)+o(\rho^{2})\right)+\rho^{\D}
\left(p_{2\Delta-d}(t)+o(\rho^{2})\right)\,,
\end{equation}
where the non-normalizable coefficient $p_{0}$ is proportional to $\lam$, while
the normalizable coefficient $p_{2\D-d}$ is proportional to $\langle \calo_{\D}
\rangle$. Similarly,
\begin{eqnarray}
A \sim \rho^{-2}\,\left( 1+ a_{d-2}(t)\rho^{d}+o(\rho^{d+4-2\Delta})
\right)\,.
  \eqlabel{ap}
\end{eqnarray}
Here, the coefficient $a_{d-2}$ controls the energy density (and pressure) of
the dual field theory, as shown in \cite{blm}. Eq.~\eqref{eoms4} is a
constraint, which in the limit $\rho\to0$, determines $\partial_t a_{d-2}$.
Integrating over $t$, we then find
\begin{eqnarray}
a_{d-2}(t) &=& {\mathcal C}- \frac{(2\D-d+1)(d-\D)}{(d-1)^2}p_{0}(t)p_{2\D-d}(t) \nonumber \\
  &&+\frac{2\Delta-d}{d-1}\int^{t}_{0}d\tilde{t}
 \ p_{2\D-d}(\tilde{t})\,\frac{d }{d\tilde{t}}p_{0}(\tilde{t})\,. \eqlabel{a24}
\end{eqnarray}
Here ${\mathcal C}=a_{d-2}(-\infty)$ is an integration constant. With $d=4$,
this expression matches that found in \cite{blmv}, using Eddington-Finkelstein
coordinates.

In our quenches, the coupling to $\calo_{\D}$ is made time-dependent with a
characteristic time $\Dt$ as
\begin{equation}
\lam=\lam\left({t}/{\Dt}\right).
\eqlabel{rate}
\end{equation}
For general $\Dt$, the response $p_{2\D-d}$ in eq.~\eqref{bc} cannot be solved
analytically.  However, as described in \cite{blm,blmv}, for large $\Dt$
(adiabatic quenches), we can find a series solution for $\phi$ in inverse
powers of $\Dt$ and in principle, we can solve for $p_{2\D-d}$ analytically.

We now present a new analytic approach for the opposite
limit of fast quenches. That is, for quenches where $\Dt$ is much smaller than
any other scale. As described above, to answer the question of how much work is
done by the quench, we need only consider the interval $0\le t\le \Dt$.
Intuitively, we may expect that when $\Dt$ is very short, there is no time for
nonlinearities in the bulk equations to become important, \ie for the metric to
backreact on the scalar.

To make this intuition manifest, we rescale the coordinates and fields by the
parameter $\Dt$ considering their (leading) dimension in units of the AdS
radius: $\rho=\Dt\,\hat{\rho}$, $t=\Dt\,\hat{t}$, $A=\hat{A}/\Dt^{2}$,
$\Sigma=\hat{\Sigma}/\Dt$ and $\phi=\Dt^{d-\D}\hat{\phi}$. With this rescaling,
the limit $\Dt\to 0$ then removes the scalar from the Einstein equations
(\ref{eoms1}--\ref{eoms4}), while leaving the form of the Klein-Gordon equation
\eqref{kg} unchanged.

The coefficient $a_{d-2}$ controls the next-to-leading order term in $A$ at
small $\rho$. As we will show, this coefficient scales as $\Dt^{d-2\D}$.
Further in eq.~\reef{ap}, this coefficient is accompanied by a factor of
$\rho^d$ and hence this term has an overall scaling of $\Dt^{2(d-\D)}$. Hence
as long as we are considering a relevant operator, this term vanishes in the limit
$\Dt\to0$. The same is true of the subleading contributions in the expression of $\Sigma$.
Hence for fast quenches with small $\Dt$, we can approximate the metric
coefficients as simply
\begin{equation}
\hat{\Sigma}=\hat{\rho}^{-1}\,,\qquad\qquad \hat{A}=\hat{\rho}^{-2}\,.
 \eqlabel{metric9}
\end{equation}
The equation for $\hat{\phi}$ becomes the Klein-Gordon equation in the AdS
vacuum spacetime, \ie
\begin{equation}
\hat{\rho} ^2 \partial^{2}_{\hat{\rho}}\hat{\phi} -(d-1) \hat{\rho}
\partial_{\hat{\rho}}\hat{\phi}-\hat{\rho} ^2 \partial^{2}_{\hat{t}}\hat{\phi}
+\D\left(d-\D\right)\hat{\phi}=0\,.
 \eqlabel{purescalar}
\end{equation}
That is, in the limit of small $\Dt$, the work done in the full nonlinear
quench can be determined by simply solving the linear scalar field equation
\reef{purescalar} in empty AdS space!

\begin{figure}[h]
\begin{center}
\psfrag{at}[c]{{$\hat{t}$}}
\psfrag{ap0}[B][tl]{{${p_0}/{\Dp}$}}
 \includegraphics[width=2in]{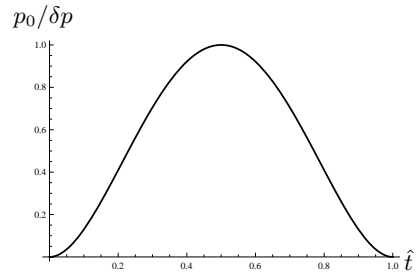}
\end{center}
  \caption{
Normalized source $p_0/\Dp$ for eq.~\eqref{sourcex} as a function of the
rescaled time $\hat{t}={t}/{\Dt}$.} \label{sourcefig}
\end{figure}

\begin{figure}[h]
\begin{center}
\psfrag{at}[c]{{$\hat{t}$}}
\psfrag{ap2}[B][tl]{{$\Dt^{2\D-d}p_{2\D-d}/\Dp$}}
 \includegraphics[width=2in]{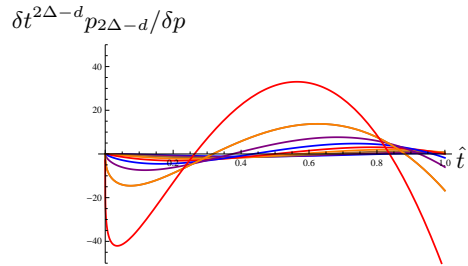}
\end{center}
  \caption{
(Colour online) The response to the source \eqref{sourcex} in $d=4$ for
$\D=2.1$ through $2.9$ in steps of $0.1$.  The plots with larger amplitudes
correspond to larger $\D$.} \label{responsefig1}
\end{figure}

\begin{figure}[h]
\begin{center}
\psfrag{at}[c]{{$\hat{t}$}}
\psfrag{ap2}[B][tl]{{$\Dt^{2\D-d}p_{2\D-d}/\Dp$}}
 \includegraphics[width=2in]{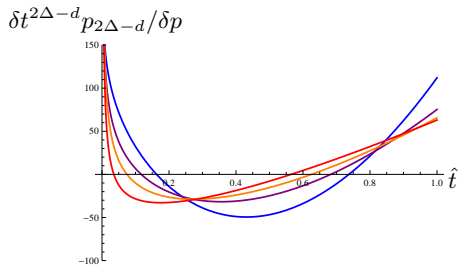}
\end{center}
  \caption{
(Colour online) The response to the source \eqref{sourcex} in $d=4$ for
$\D=3.1$ through $3.4$ in steps of $0.1$.  The colours blue, purple, orange,
and red correspond to the response for $\D=3.1$ through $3.4$
respectively.} \label{responsefig2}
\end{figure}

Now we consider sources that vanish for $t\le 0$ and are constant for
$t\ge\Dt$. In $0<t<\Dt$, we vary the source as
\begin{equation} \eqlabel{source}
p_{0} (t) =  \Dp\ \left({t}/{\Dt}\right)^{\kappa}
\end{equation}
where $\kappa$ is a positive exponent. Note that here $p_{0}(t\ge\Dt)=\Dp$.
Since $\phi=0$ before we switch on the source at $t=0$, it remains zero
throughout the bulk up to the null ray $t=\rho$. Therefore we impose
\begin{equation} \eqlabel{bdy}
\phi(t=\rho,\rho) = 0\,.
\end{equation}
Evaluating the scalar field equation \reef{purescalar} subject to the boundary
conditions \eqref{source} and \eqref{bdy}, we find \cite{prep}
\begin{equation} \eqlabel{p2}
p_{2\D-d}(t) = b_\kappa\ \Dt^{d-2\D}\ \Dp\ \left(t/{\Dt}\right)^{d-2\D+\kappa}
\end{equation}
with
\begin{equation}
b_\kappa=-\frac{2^{d-2\D}\,\Gamma(\kappa+1)\, \Gamma(\frac{d+2}{2}-\D)}
{\Gamma(d+1+\kappa-2\D)\,\Gamma(\D-\frac{d-2}{2})}\ .
\eqlabel{bn}
\end{equation}
Of course, if we construct more complicated sources with a series expansion of
monomials as in eq.~\reef{source}, then since eq.~\reef{purescalar} is linear,
the response is simply given by the sum of corresponding terms as in
eq.~\reef{p2}. As an example, consider the source
\begin{equation} \eqlabel{sourcex}
p_0(\hat{t})=16\, \Dp \left(\hat{t}^2-2\hat{t}^3+\hat{t}^4 \right)\,
\end{equation}
as shown in fig.~\ref{sourcefig}. In this case, the source vanishes in both the
initial and final state and it reaches the maximum $\Dp$ at $t=\Dt/2$.
Figs.~\ref{responsefig1} and \ref{responsefig2} show the corresponding response
for various values of $\D$ in $d=4$.

The response coefficient \reef{p2} exhibits two noteworthy features: First, we
see that the overall scaling of the response is $\Dt^{d-2\D}$. This is
precisely the behaviour found in the numerical studies of \cite{blmv} in the
case $d=4$. Second of all, $p_{2\D-d}$ varies in time as $t^{d+\kappa-2\D}$.
Therefore if $\kappa<2\D-d$, the response (\ie the operator expectation value
$\langle\mathcal{O}_{\D}\rangle$ in the boundary theory) diverges at $t=0$! For
a source constructed as a series, both of these features in the response are
controlled by the smallest exponent, as illustrated in figs.~\ref{responsefig1}
and \ref{responsefig2} for eq.~\reef{sourcex}.

For homogeneous quenches, the diffeomorphism Ward identity reduces to
$\partial_t \mathcal{E}=- \langle{\cal O}_\Delta\rangle\,\partial_t \lambda$
\cite{blm,blmv}. Hence we can evaluate change in the energy density as
\begin{equation} \eqlabel{energy}
\D\mathcal{E}=-\cala_\cale \int^{+\infty}_{-\infty}p_{2\D-d}\,\partial_t p_{0}\,dt\,,
\end{equation}
with \cite{ctdef}
\begin{equation} \eqlabel{aenergy}
\cala_\cale=\frac{2\D-d}{16\pi G_{d+1}}
= \frac{(2\D-d)\pi^{d/2}\, \Gamma\!\left(\frac
d2\right)}{2d(d+1)\,\Gamma(d-1)}\ C_T \,.
\end{equation}
Since $\partial_t p_{0}$ vanishes for $t<0$ and $t>\Dt$, the above integral
reduces to an integral from $0$ to $\Dt$. It is for this reason that we do not
need to determine the response $p_{2\D-d}$ after $t=\Dt$. Further, for fast
quenches, the change in energy density will scale as $\Dt^{d-2\D}$. Note that
$\partial_t p_{0}$ scales as $\Dt^{-1}$, but the range of the integral
$0<t<\Dt$ adds an additional scaling of $\Dt^{+1}$. Hence the net scaling of
$\D\mathcal{E}$ is precisely the scaling of $p_{2\D-d}$. Again this precisely
matches the scaling found numerically in \cite{blmv} for $d=4$. In fact, this
behavior can be fixed as follows: Since eq.~\reef{purescalar} is linear, we
must have $p_{2\D-d}\propto\Dp$ and hence $\D\mathcal{E}\propto \Dp^2$ from
eq.~\reef{energy}. Finally, dimensional analysis demands $\D\mathcal{E}\simeq
\Dp^2/\Dt^{2\D-d}$, up to numerical factors.

However, recall the singular behaviour in the response at $t=0$ for
$\kappa<2\D-d$. Despite this divergence, one can easily see that in fact, the
corresponding integral \reef{energy} remains finite as long as
$\kappa>\D-\frac{d}{2}$. That is, for fixed $\D$ and $d$, we are constrained as
to how quickly the source may be turned on. In fact, a more careful examination
\cite{prep} of the bulk solutions indicates that our analysis is valid for
$\kappa>\D-\frac{d}{2}+\frac{1}2$. For quenches not satisfying this inequality,
we can no longer ignore the backreaction of the scalar on the spacetime
geometry.

To summarize, we have showed that in the limit of fast, abrupt quenches, the
response and the energy density of a strongly coupled system which admits a
dual gravitational description scales as $\Dt^{d-2\D}$. Here $\frac{d}{2}\le \D
<d$ is the conformal dimension of the  quenched operator in the vicinity of the
ultraviolet fixed point. Although we considered a quench from a vacuum state at
$t=0$, our results are universal. That is, they are independent of the initial
state of the system, \eg we may start with a thermal state, as in
\cite{blm,blmv}. This is again a reflection of the fact that abrupt holographic
quenches are completely determined by the UV dynamics of the theory
--- see fig.~\ref{triangle}. Also, if different operators are quenched
simultaneously, the response is dominated by the one with the largest conformal
dimension.

We emphasize that while our calculations only considered the linearized scalar
equation \reef{purescalar}, our results apply for the full nonlinear quench. In
the limit $\Dt\to 0$, the relevant physics occurs in the far asymptotic
geometry (see fig.~\ref{triangle}) where the bulk scalar and perturbations of
the AdS metric are all small. This contrasts with \cite{blm,blmv}, which only
worked within a perturbative expansion in the amplitude of the scalar. Of
course, the scalings determined there match those found here, but it was
uncertain if they would persist in a full nonlinear analysis.

Of course, the present analysis does not  predict the dynamical evolution of
the system for $t> \Dt$, however, we can deduce the equilibrium thermal state
of the system as $t\to \infty$. Indeed, since the coupling and energy density
are constant for $t> \Dt$, $\lambda(+\infty)=\lambda(\Dt)$ while
eq.~\reef{energy} determines the final energy density of the system, to leading
order in $\delta t$. Together, these parameters completely specify the final
equilibrium state.

Note that our analysis strictly applies to relevant operators, for which
$d-\Delta>0$. With a marginal operator (\ie $\Delta=d$), we can expect
$\D\mathcal{E}\propto\Dt^{-d}$ on purely dimensional grounds \cite{cy1}. While
this matches the scaling found above, our numerical coefficients would no
longer be valid. Marginal operators were also considered in \cite{early5,new2}
with a four-dimensional bulk. This case is analytically accessible because the
scalar propagates on the light-cone. Extending this analysis to an
odd-dimensional bulk is more challenging \cite{early5} because the scalar
propagator is nonvanishing throughout the interior of the light-cone, similar
to that for the relevant operators studied here.

Our discussion was also limited to $\frac{d}2\le \D<d$, while unitarity bounds
also allow for $\frac{d}2-1\le \D<\frac{d}2$. In the latter range, we must
consider the so-called `alternate quantization' of the bulk scalar
\cite{alternate}. In fact, the asymptotic expansion of the scalar takes
precisely the same form as in eq.~\reef{bc}. However, in this regime, $p_{0}$
($p_{2\D-d}$) is the coefficient of the (non-)normalizable mode. Our analysis
applies equally well for this range of $\D$ and so one still finds
$p_{2\D-d}\simeq\Dp\,\Dt^{d-2\D}$. That is, the response becomes vanishingly
small as $\Dt\to0$ with $\Dp$ kept fixed. Hence to produce a finite
$\langle{\cal O}_\D\rangle$ or finite $\D\mathcal{E}$, we would need to scale
$\Dp$ with an inverse power of $\Dt$.

When $\Delta$ is an integer for even $d$ or half-integer for odd $d$, the
scaling of the response $\langle {\cal O}_\D\rangle$ receives additional
$\log({\Dt})$ corrections \cite{blm}. These logarithmic corrections arise from
$\log\rho$ modifications in the asymptotic expansion \eqref{bc} of the bulk
scalar and are easily computed analytically following the present approach
\cite{prep}.

Another exceptional case arises with $\kappa=2\Delta-d-n$ where $n$ is a
positive integer. In this case, eq.~\reef{bn} indicates $b_\kappa=0$. Hence if
the source is given by a series of monomials \reef{source}, the scaling of the
response will be controlled by the first subleading contribution. With a single
monomial, the (subleading) scaling of the response is controlled by
nonlinearities in the bulk equations \cite{prep}, \ie $p_{2\D-d}\simeq
\Dt^{-\D} (\Dp\, \Dt^{d-\D})^n$ where $n=2$ if the potential contains a
$\phi^3$ term and $n=3$ otherwise.

It is interesting to consider the limit of abrupt quenches with $\Dt=0$, as
this usually sets the starting point in analyses at weakly coupling. Our
holographic result, $\D\mathcal{E}\simeq \Dp^2/\Dt^{2\D-d}$, indicates that the
energy density diverges for an abrupt quench with $\D>\frac{d}2$ (a logarithmic
divergence appears for $\D=\frac{d}2$ \cite{blm,prep}). Hence it would be
interesting to carefully compare these holographic results with those for the
weak coupling calculations of, \eg \cite{cc2,cc4,new1}. Let us note here that
certain singular behaviours were observed for abrupt quenches of a fermionic
mass term \cite{new1}. Of course, the preceding considerations assume $\Dp$ is
held fixed in the limit $\Dt\to0$. Instead, if we scale the source to zero as
$\Dp\propto\Dt^{\D-\frac{d}2}$, $\D\mathcal{E}$ will remain finite. However, we
stress that this limit still produces a divergent response since
$p_{2\D-d}\sim\Dt^{d-2\D}\,\Dp \propto\Dt^{\frac{d}2-\D}$.

An important question to ask is to what extent our results are relevant for
everyday physical systems. Gauge theories with a dual gravitation description
are necessarily strongly coupled and have an ultraviolet fixed point with large
central charge. The framework of the gauge-string duality allows for the study
of both the finite 't Hooft coupling corrections (the higher-derivative
corrections in the gravitational dual) and non-planar (quantum string-loop)
corrections. We expect that our gravitational analysis are robust with respect
to the former, as the relevant near-boundary space-time region is weakly
curved. Whether finite central charge corrections are important or not is an
open question.

\noindent {\bf Acknowledgements:} We would like to thank David Berenstein, Luis
Lehner, David Mateos, Shiraz Minwalla, Jo\~{a}o Penedones, Misha Smolkin and
Julian Sonner for useful discussions. AB and RCM would like to thank Lorentz
Center for hospitality, where part of this work was done. Research at Perimeter
Institute is supported by the Government of Canada through Industry Canada and
by the Province of Ontario through the Ministry of Research \& Innovation. AB
and RCM gratefully acknowledge support from NSERC Discovery grants. Research by
RCM is further supported by funding from the Canadian Institute for Advanced
Research.


\begin{thebibliography}{99}
\bibitem{ctdef} $C_T$ is a `central charge' characterizing the leading
    singularity in the two-point function of the stress tensors --- see
    \cite{cc} for details.
\end{thebibliography}
\end{document}